\documentclass[twocolumn]{aastex7} 

\begin{document}

\title{Multi-fluid multi-species models for inverse FIP-effect}

\author[0000-0002-0333-5717]{Juan Martinez-Sykora}
  \affiliation{Lockheed Martin Solar and Astrophysics Laboratory, 3251 Hanover Street, Palo Alto, CA 94306, USA}
  \affiliation{SETI Institute, 339 Bernardo Ave, Mountain View, CA 94043, USA}
    \affil{Rosseland Centre for Solar Physics, University of Oslo, P.O. Box 1029 Blindern, NO–0315 Oslo, Norway}
	\affil{Institute of Theoretical Astrophysics, University of Oslo, P.O. Box 1029 Blindern, NO–0315 Oslo, Norway}
 \email{jmsykora@seti.org}

\author[0000-0002-6597-4415]{Paola Testa}
	\affiliation{Harvard-Smithsonian Center for Astrophysics, 60 Garden Street, Cambridge, MA 02193, USA}
 \email{ptesta@cfa.harvard.edu}

\author[]{Deborah Baker}
	\affiliation{Mullard Space Science Laboratory, University College London, Holmbury St. Mary, Dorking, Surrey, RH5 6NT, UK}
    \email{deborah.baker@ucl.ac.uk}

 \author[0000-0002-8370-952X]{Bart De Pontieu}
	\affiliation{Lockheed Martin Solar and Astrophysics Laboratory, 3251 Hanover Street, Palo Alto, CA 94306, USA}
	\affil{Rosseland Centre for Solar Physics, University of Oslo, P.O. Box 1029 Blindern, NO–0315 Oslo, Norway}
	\affil{Institute of Theoretical Astrophysics, University of Oslo, P.O. Box 1029 Blindern, NO–0315 Oslo, Norway}
    \email{bdp@lmsal.com}

\begin{abstract}
The inverse First Ionization Potential (FIP) effect is rarely observed in the solar atmosphere, and this anomaly poses a challenging problem in understanding physical processes driving this chemical fractionation. In this work, we investigate various scenarios where the inverse FIP effect could occur using simplified 1D multi-fluid MHD models. The model treats the full MHD equations with multi-fluid and multi-species effects, rather than using wave analysis to derive the ponderomotive force and semi-empirical 1D models. In the parametric study considered here, for upward Alfvén waves, one can achieve a negative (opposite) ponderomotive force when the magnetic field strength and the magnetic flux tubes' expansion with height counteract the dissipation and damping effects from multi-fluid interactions.
\end{abstract}

\section{Introduction}

Abundances vary across the solar atmosphere and stellar types (e.g.,\cite{Testa:2010fk,Laming:2015cr,Testa:2015xi}). The abundance variation in the solar corona and wind has been strongly connected to the slow and fast solar wind \cite{geiss:1995SSRv...72...49G}, the magnetic field topology, and upflows \cite{Brooks:2011mi}. Therefore, understanding the mechanisms that change abundances will provide insight into the drivers of solar/stellar activity and the solar/stellar winds.

For later spectral types (G K M), the dependence of the abundances on the First Ionization Potential (FIP) decreases, and at about K5, an inverse FIP (iFIP) effect is often observed \cite{Wood2010ApJ...717.1279W,Seli2022AA...659A...3S}. High magnetic activity (high X-ray emission) stars, such as the multiple system HR 1099, also typically have abundances showing a strong inverse correlation with the FIP \cite{Brinkman2001AA...365L.324B,Testa:2015xi}.

The inverse FIP effect is rarely observed in the solar atmosphere. Still, there are certain locations where the iFIP has been observed. These localized regions are typically in an active region (AR) with complex magnetic field topology, with strong magnetic fields, such as strong light bridges in sunspot umbrae of coalescing or merging sunspots, or transient moments during flaring events (see \cite{Baker2024ApJ...970...39B}, and references therein; also \cite{Baker_2026} in this Topical Issue) as shown with Hinode/EIS\cite{Culhane:2007fk}. In \cite{Baker_2026}, IRIS\cite{De-Pontieu:2014yu} observations are used to show that the iFIP effect occurs in association with increased non-thermal broadening in the chromosphere. Being able to map those locations on the Sun should also help us to understand the iFIP effect in other stars. Hence, it is of great interest to understand chemical fractionation in the Sun and other stars.

Since the variation (relative to the photosphere) of solar and stellar abundances in the wind is typically correlated or anti-correlated to the FIP of the elements, and those elements become ionized in the solar chromosphere, the chemical fractionation must occur somewhere in the chromosphere. This region is highly complex due to the many physical processes playing a role: radiative losses, non-LTE effects, non-equilibrium (NEQ) ionization, and partial ionization effects occurring simultaneously, along with numerous transitions with height: from non-magnetized to magnetized, from collisional to collisionless, and from gas pressure-dominated to magnetic pressure. Hence, to address chemical fractionation, these effects are important to consider. However, until now, the state-of-the-art and most popular model is based on semi-empirical 1D models of the derived ponderomotive force from imposed Alfvén waves to address this problem \cite{Laming:2004qp}. Such models, by necessity, ignore many of the complex interactions that occur in the chromosphere.

In this work, we will investigate, for the first time using a self-consistent multi-fluid approach and for highly simplified scenarios, what would be required to have a ponderomotive force that could lead to an iFIP abundance dependence. For this, we use the multi-fluid, multi-species (MFMS) 3D MHD numerical code (Ebysus, \cite{Martinez-Sykora:2020ApJ...889...95M}), as briefly described in the following section. Results and analysis of the parametric study of different numerical simulations of Alfvén waves are detailed in Section~\ref{sec:res}. The discussion and conclusions are detailed in Section~\ref{sec:dis}. 

\section{Ebysus: a multi-fluid and multi-species numerical code}~\label{sec:eby}

The MFMS numerical code Ebysus (e.g., \cite{Martinez-Sykora:2020ApJ...900..101M,Evans2023ApJ...949...59E}) allows solving the MHD equations as a separate fluid for each excited and/or ionized state of each considered species. This code is modular and allows for the inclusion of ionization/excitation in NEQ, collisions and interaction between different fluids, thermal conduction, radiative losses, and the Hall term (see \cite{Khomenko:2014nr, Ballester:2018fj} for multi-fluid and multi-species derivation). Recently, we have included in this code the second-order Partitioned Implicit-Explicit Orthogonal Runge-Kutta (PIROCK) method, allowing for advancing in time by combining efficient explicit stabilized and implicit integration techniques while employing variable time-stepping with error control \cite{Wargnier2025AA...695A.262W}.

For the present study, following a similar setup as in \cite{Martinez-Sykora:2020ApJ...900..101M}, we simplify the problem by not including ionization or recombination, thermal conduction along the magnetic field lines, or gravity. Similar to \cite{Martinez-Sykora2023ApJ...949..112M}, we assumed five species (hydrogen, helium, iron, silicon, and calcium, with two ionization states (ground and first ionization) as shown in table~\ref{tab:pop} with initial uniform densities and a temperature of 16000~K. For simplicity, in this work, we considered a fixed density; note that the electron number density within the chromosphere can vary within $\sim10^{9}-10^{13}$ cm$^{-3}$ in chromospheric AR (e.g., \cite{Testa2023ApJ...944..117T}). In the future, this variation needs to be addressed. The abundances considered are photospheric \cite{Asplund:2009uq}.

\begin{figure}[!ht]
\centering\includegraphics[width=3.2in]{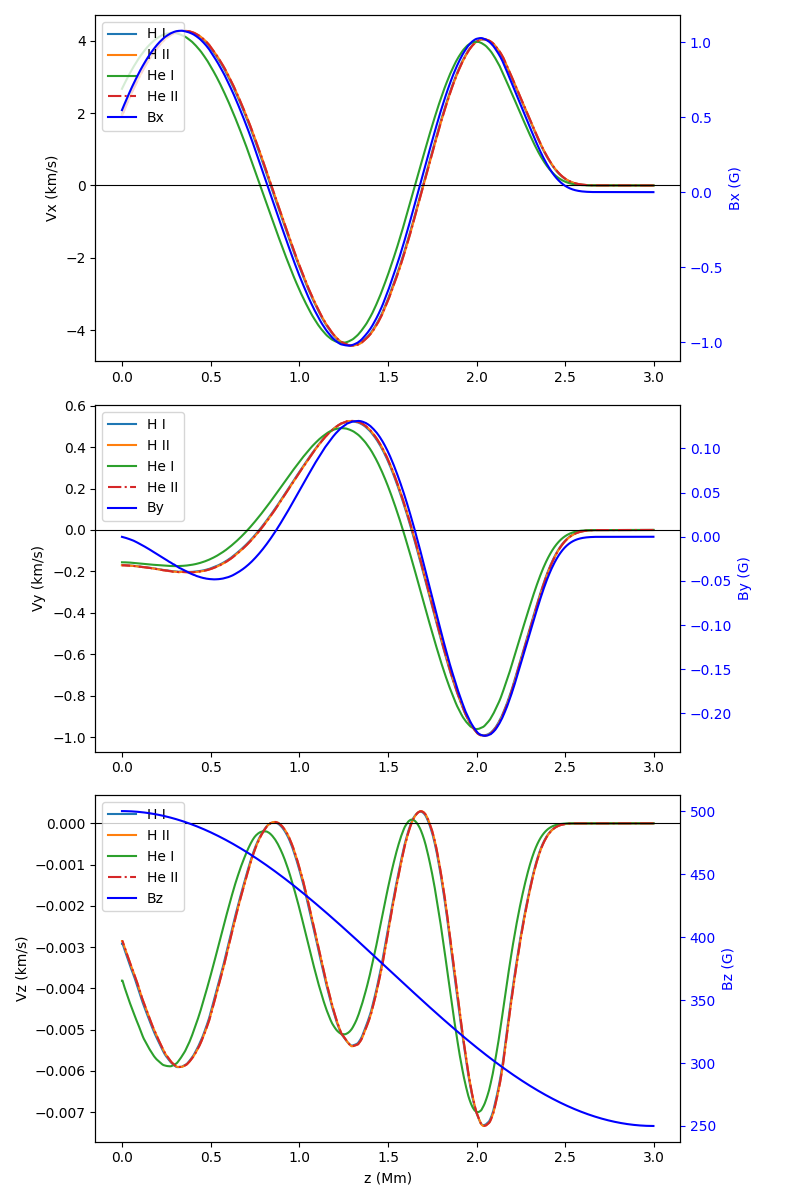}
\caption{Velocity for the hydrogen and helium species and magnetic field variation as a function of height at $t=1.5$~s (x, y, z, respectively, from top to bottom panels). For simplicity, we limit the number of field lines to the two dominant species, i.e., H and He for the simulation B0502.}
\label{fig:strat}
\end{figure}

The 1-dimensional numerical domain is 3~Mm long with 3~km grid spacing as shown in figure~\ref{fig:strat}. The collisions include resonant absorption for hydrogen-hydrogen and hydrogen-helium species interactions, otherwise molecular Maxwell collisions. We also consider Coulomb collisions for charged species (see \cite{Wargnier:2022ApJ...933..205W} for details), although they are usually negligible. \vspace*{-6pt}

\begin{table}[!ht]
	\caption{Density of the various fluids. From left to right: the species, and the number density for the neutral and ionized fluid in cm$^{-3}$}\label{tab:pop}	
	\begin{tabular}{|c|c|c|}
		\hline
		Species & Neutrals & Ions  \\ \hline  \hline
		H &  $10^{11.1}$  & $10^{10.71}$\\ \hline
        He & $10^{10.4}$ & $10^{7.8}$\\ \hline
		Fe & $10^{4.5}$ & $10^{5.6}$ \\ \hline
		Si & $10^{4.8}$ & $10^{5.7}$ \\ \hline
		Ca & $10^{2.5}$ & $10^{5.4}$ \\ \hline
	\end{tabular}
\vspace*{-4pt}
\end{table}

We run a parameter range where we vary the expansion of the magnetic flux tube with height and the field strength to investigate in which scenarios we obtain iFIP, as shown in table~\ref{tab:1ds}. Note, the way we define the expansion is $\frac{- \Delta B_z}{<|B_z|>\, \Delta z} > 0$, i.e., cases with the same expansion (B2010, B1005 and B0502) will have same $\frac{- \Delta B_z}{<|B_z|>\, \Delta z}$. Note that, in this 1.5D domain, the divergence of the magnetic field is not zero. Similarly, we only vary the $B_x$ component at the boundary \cite{Martinez-Sykora2023ApJ...949..112M}.\vspace*{-6pt} 

\begin{table*}[!ht]
	\caption{List of numerical simulations. From left to right: the name, magnetic field configuration, Alfvén speed range, and frequency and amplitude of the Alfvén wave driver. In the second column, the first value of the $B$ range is the bottom boundary, and the second is the top, so $[2000, 1000]$~G is a loop whose magnetic field strength decreases with $z$, i.e., direction of wave propagation.}\label{tab:1ds}	
	\begin{tabular}{|c|c|c|c|c|c|c|c|}
		\hline
		Name & $B$ range (G) & $v_A$ (km~s$^{-1}$) & Fr. (Hz) & $B$ Amp. wave (G) \\ \hline  \hline
		B2015 & $[2000,1500]$ & $[19237,14427]$ & 1   & 0.1 \\ \hline
		B2010 & $[2000,1000]$ & $[19237,9618]$ & 1   & 0.1 \\ \hline
		B2005 & $[2000,500]$ & $[19237,4809]$ & 1   & 0.1 \\ \hline
		B1005 & $[1000,500]$ & $[9618,4809]$ & 1   & 0.1 \\ \hline
		B0502 & $[500,250]$ & $[4809,2404]$ & 1   & 0.1 \\ \hline
	\end{tabular}
\vspace*{-4pt}
\end{table*}

The ``top" boundary is open. At the ``bottom", we drive an Alfvén wave along the component $x$ of the magnetic field with a frequency of 1~Hz and 0.1~G amplitude. Similar to \cite{Martinez-Sykora2023ApJ...949..112M}, we use quotation marks around the words ``top" and ``bottom" because our numerical experiments do not include gravity. The bottom-to-top direction is the direction of wave propagation.

\section{Results}~\label{sec:res}

In \cite{Martinez-Sykora2023ApJ...949..112M}, the presence of the various fluids leads to a collisional and, for the ionized fluids, an electric coupling. These couplings introduce an offset in the wave-related dynamics of the various fluids and damp Alfvén waves as they propagate through the atmosphere. This coupling and damping lead to a positive ponderomotive acceleration, i.e., in the direction of the propagating wave. If the wave propagates upward, the ponderomotive force is upward, and for downward-propagating waves, the ponderomotive direction is downwards too. So, to cause a FIP enhancement of abundances in the upper part of the domain, waves will propagate upward from the chromosphere under this parametric study. 

Considering the parameter range in \cite{Martinez-Sykora2023ApJ...949..112M}, we notice that the stronger the expansion of flux tubes with height, the smaller the ponderomotive force. So, one could imagine that with sufficient expansion of the flux tube (i.e., decrease of the magnetic field strength with height), the ponderomotive force might reverse its direction. In addition, one aspect not included in that work was the dependence of the abundance enhancement on the magnetic field strength. Our goal with the simulations listed in Table~\ref{tab:1ds} is to investigate these two scenarios.

\begin{figure}[!ht]
\centering\includegraphics[width=3.2in]{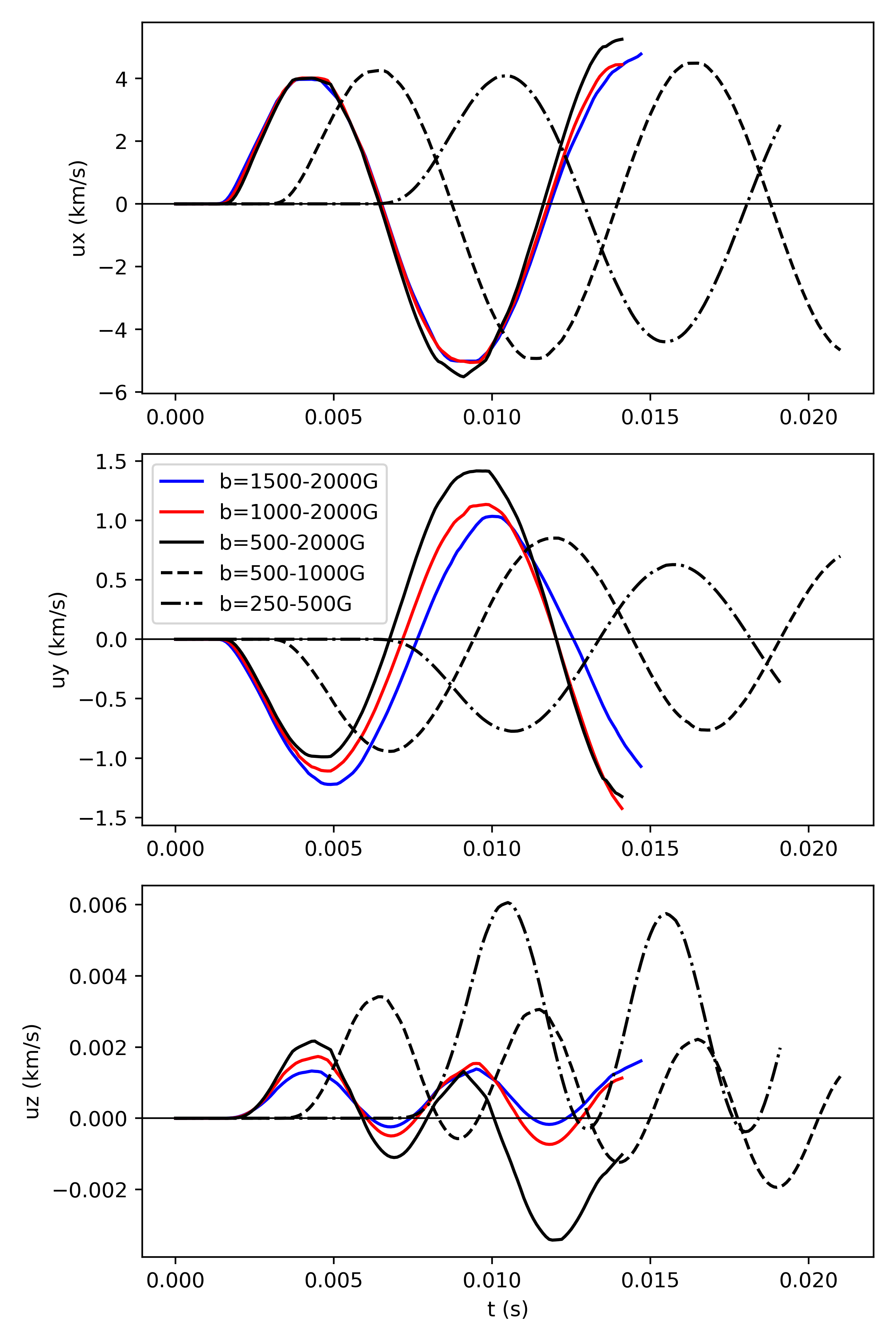}
\caption{Time evolution at $z=1.5$~Mm of the velocity components (x, y, z, respectively, from top to bottom panels) for the proton fluid. Different colors show different cases for magnetic field strength decrease with height, and different line styles show, for the same magnetic field expansion factor, different field strengths as labeled in the middle panel.}
\label{fig:time-evo}
\end{figure}

Figure~\ref{fig:time-evo} summarizes the most relevant properties across the parameter range study listed in table \ref{tab:1ds}. The first thing to notice is that, due to the different Alfvén speeds for each considered case (Table~\ref{tab:1ds}, the stronger the magnetic field (different line-styles), the sooner the Alfvén wave reaches the same height ($z=1.5$~Mm). For the weakest magnetic field case (dashed-dotted line) this occurs around $t=0.006$~s, for the second weakest field case (dashed) around $t=0.004$~s and for the strongest case around $t=0.002$~s. Similarly, the wavelength increases with field strength, not shown here. Like in \cite{Martinez-Sykora:2020ApJ...900..101M} and \cite{Martinez-Sykora2023ApJ...949..112M}, we find that although the driver is along the x-axis, the magnetic field and the ion motions rotate into the y direction. This can be seen in the middle panel as a temporal offset. This occurs because of the coupling between the various ionized fluids and the Hall term. This offset between the x and y components of the velocity changes for different values of the decrease of the magnetic field with height (simulations B2005, B2010, and B2015 shown with different colors in the middle panel of Figure~\ref{fig:time-evo}). It is also interesting to see that the amplitude of the y component of the velocity decreases with field strength  (simulations B2010, B1005, and B0502 shown with different line styles in the middle panel of Figure~\ref{fig:time-evo}). 

The focus of this study is the impact of the loop's magnetic field properties on the ponderomotive force, i.e., the acceleration of ions along the loop. Under the conditions and within the parametric study performed here, it is very interesting to see, in the bottom panel (which shows the velocity along the z direction), that a strong decrease of the magnetic field (i.e., expansion of flux tube) with height can indeed have the opposite sign on the direction of the acceleration of the ions along the loop. Note how the more strongly the field strength decreases with height, the more negative the z component of the velocity becomes (simulations B2005, B2010, and B2015 shown with different colors in Figure~\ref{fig:time-evo}). The variation of the Alfvén wave amplitude with height is the result of a competition between the expansion of the magnetic field and the dissipation due to fluid interactions. This variation will lead to a ponderomotive force along the loop. Note that in \cite{Martinez-Sykora2023ApJ...949..112M} the wave dissipation due to fluid interactions dominates and leads to a positive (upward) ponderomotive force with an upward propagating Alfvén wave for the parameter range considered in that study (see panel g of figure 13 in \cite{Martinez-Sykora2023ApJ...949..112M}). However, here, where we impose stronger flux tube expansion and stronger field strength, the expansion of the magnetic field dominates the variation of the Alfvén waves compared to the wave dissipation due to fluid interactions. Note that this acceleration is different and independent from the acceleration caused by the impact of the variation with height of the flux tube cross-section in the advection term in a 1D approach that describes a flux tube that expands with height \cite{Hollweg1982SoPh...75...35H}.

The magnetic field strength also plays a role in the slope. For the same field expansion factor, the stronger the magnetic field, the smaller the amplitude along the z-axis, and the more negative the slope is with time (simulations B2010, B1005, and B0502 shown with different line styles in Figure~\ref{fig:time-evo}). This is expected since with a stronger magnetic field, stronger the Lorentz force, larger the Alfvén speed, and the shorter the travel time of the wave through the chromosphere which is where the dissipation could play a role. 

\section{Discussion and Conclusion}~\label{sec:dis}

One of the most popular models to explain the iFIP effect is based on the effects of the ponderomotive force, using semi-empirical models (e.g., \cite{Laming:2004qp}). In this study, we simulate Alfv\'enic waves using a full multifluid MHD treatment. Hence, we go beyond the limitations of using semi-empirical models and assumptions on the derivation of the ponderomotive force. Still, we consider highly simplified scenarios to investigate what might be required to achieve iFIP within the parametric study considered here, including multifluid effects.

We computed multifluid numerical models of Alfvén waves in a nonuniform magnetic field configuration. One must bear in mind that the multifluid models used here are highly simplified: they do not include density stratification or gravity, nor do they capture chromospheric dynamics, ionization/recombination, thermal conduction, or wave reflections. 

As shown in \cite{Martinez-Sykora2023ApJ...949..112M}, the presence of different fluids (ionized and neutral fluids for various species) leads to both collisional and electric coupling, damping the Alfvén waves while traveling through the atmosphere. This damping leads to a positive ponderomotive acceleration, i.e., in the direction of the propagating wave. In the present work, we have expanded the parametric study to aim for a negative ponderomotive acceleration, i.e., in the opposite direction of the propagating wave. This change in the direction of the ponderomotive acceleration would lead to an iFIP effect for a chromospheric driver. We succeeded in producing the opposite acceleration to the propagating waves by having a stronger average field strength and a steeper decrease of the field strength with height. In other words, we have stronger flux tubes that expand more with height. These two properties can counteract the impact of collisions in the partially ionized chromosphere on the damping of Alfv\'en waves and can change the direction of the ponderomotive force.

We limited the study to a single high frequency case (1Hz) and single and fixed density. However, as seen in \cite{Martinez-Sykora2023ApJ...949..112M}, the lower frequency waves will decrease the ponderomotive force. In addition, increasing the amplitude of the wave could increase the role of the dissipation and increase the ponderomotive force (more positive). Similarly, varying the density will change the coupling across the different species and the damping of the Alvén waves, hence changing the ponderomotive force. However, further studies are required to address this under the parametric study considered here. 

The need for a strong field and strong flux tube expansion with height agrees with the observational finding that, in the solar atmosphere, the iFIP effect is seen around lightbridges in strong, complex sunspots \cite{Baker2024ApJ...970...39B}. It is not unreasonable to assume that the magnetic field in such regions is strong and expected to decrease rapidly with height.  This result can also help to interpret stellar observations where the iFIP effect is often seen for highly active stars and later spectral type stars such as M-dwarfs. These stars are expected to have strong magnetic fields, which decrease rapidly with radial distance from the surface. Based on our simulations, the ponderomotive force would be negative for Alfvén waves propagating outward, producing the iFIP effect.

The scenario presented in this work is an alternative explanation for iFIP compared to the work by \cite{Laming2021ApJ...909...17L}. \cite{Laming2021ApJ...909...17L} postulates that upward propagating fast mode waves, generated by subsurface reconnection, are refracted downwards in the chromosphere and cause downwards acceleration of ions due to the ponderomotive force. \cite{Laming2021ApJ...909...17L} ignored collisional time-scales among other ad-hoc assumptions, e.g., subphotospheric reconnection. Our models do not require subsurface reconnection and naturally drive the iFIP effect under conditions with strongly expanding flux tubes, which are likely to occur in light bridges \cite{Baker_2026}. Note that coordinated observations of the dynamics in the chromosphere, e.g., with IRIS, and the abundances in the corona, e.g., with Hinode/EIS, are critical to constrain these models. For example, observations of reconnection, e.g., by studying strong flows associated with UV bursts, or the dependence with height of non-thermal velocities (e.g., due to damping using IRIS$^{2+}$\cite{SainzDalda2024ApJS..271...24S} would be very useful to constrain the models.

Another scenario to produce iFIP would be based on the work by \cite{Martinez-Sykora2023ApJ...949..112M}, but in that case, the driver must be in the corona, and the wave propagates downwards. If the collisions damp the Alfvén wave, the ponderomotive force would point downward. This would enable the iFIP effect, for example, during strong flares when reconnection in the corona generates strong and downward-propagating Alfv\'en waves. In this case, the damping would need to dominate the field expansion in the direction of the propagating Alfvén wave. This seems reasonable since, in most cases, the magnetic field lines expand from the lower to the upper atmosphere.

\enlargethispage{20pt}

\section{Acknowledgements} 
We gratefully acknowledge support by NASA contract NNG09FA40C (IRIS) and 80GSFC21C0011 (MUSE), NASA grant 80NSSC26K0018, and NSF grants AGS2532363 and AGS2532187. The simulations have been run on clusters, the Pleiades cluster, through the computing projects s1061 and s2601 from the High-End Computing (HEC) division of NASA. Hinode is a Japanese mission developed and launched by ISAS/JAXA, collaborating with NAOJ as a domestic partner, and NASA and STFC (UK) as international partners. Scientific operation of Hinode is performed by the Hinode science team organized at ISAS/JAXA. This team mainly consists of scientists from institutes in the partner countries. Support for the post-launch operation is provided by JAXA and NAOJ (Japan), STFC (UK), NASA, ESA, and NSC (Norway). D.B. is funded under Solar Orbiter EUI Operations grant number ST/X002012/1 and Hinode Ops Continuation 2022-25 grant  number ST/X002063/1. We thank the Royal Society for its funding and support throughout the organization of the Theo Murphy meeting Solar Abundances in Space and Time and production of this special issue.


\bibliographystyle{aasjournal}
\bibliography{collectionbib.bib}

\end{document}